\documentclass[11pt,amssymb,amsmath]{article}

\def\la{\langle}
\def\ra{\rangle}
\def\a{\alpha}
\def\b{\beta}

\def\h{\hskip 1cm}

\usepackage{graphicx,psfrag,bm,amsmath}
\begin{document}
\begin{titlepage}
\vspace{4cm}
\begin{center}{\Large \bf Thermal effects on quantum communication through spin chains}\\
\vspace{1cm}A.Bayat\footnote{email:${\rm abolfazl\_}{\rm
bayat}$@mehr.sharif.edu},
\hspace{0.5cm} V. Karimipour \footnote{Corresponding author, email:vahid@sharif.edu}\\
\vspace{1cm} Department of Physics, Sharif University of Technology,\\
P.O. Box 11365-9161,\\ Tehran, Iran
\end{center}
\vskip 3cm
\begin{abstract}
We study the effect of thermal fluctuations in a recently proposed
protocol for transmission of unknown quantum states through
quantum spin chains. We develop a low temperature expansion for
general spin chains. We then apply this formalism to study exactly
thermal effects on short spin chains of four spins. We show that
optimal times for extraction of output states are almost
independent of the temperature which lowers only the fidelity of
the channel. Moreover we show that thermal effects are smaller in
the anti-ferromagnetic chains than the ferromagnetic ones.
\end{abstract}
\end{titlepage}

\vskip 3cm
\section{Introduction}\label{intro}
One of the basic ingredients of quantum communication is the
transport of a known or unknown quantum state from one point to
another \cite{benn}. Recently it has been shown that quantum spin
chains can act as channels for the efficient \cite{bose,kor,song}
or perfect \cite{bose2,shi} transfer of quantum states. Of
particular interest to us here is the works reported in
\cite{bose,kor}. It has been shown that this scheme may be used
for linking several small quantum processers in large scale
quantum computing. In this scheme, an $N$-site ferromagnetic
Heisenberg spin chain placed in a magnetic field, plays the role
of the quantum channel. The spins of the channel are numbered from
$1$ to $N$. It is assumed that this chain is in its ground state,
$|-,-,\cdots -\ra$, where $|-\ra$ denotes the state of a spin in
the negative $z$ direction and the magnetic field points to the
positive $z$ direction. One then adds a spin or qubit to the left
hand side of this channel labeled by zero, which is in an unknown
state $|\phi\ra $ (figure 1-a). The state of this qubit is to be
transmitted with a high fidelity to the right hand
side, by the natural time evolution of the chain.\\
In this way one may circumvent a problem which exists in quantum computer implementation namely the difficulty
of switching on and off between spins\cite{zhou,benjamin}.\\

\begin{figure}\label{mychain2}
\centering
    \includegraphics[width=10cm,height=20cm,angle=0]{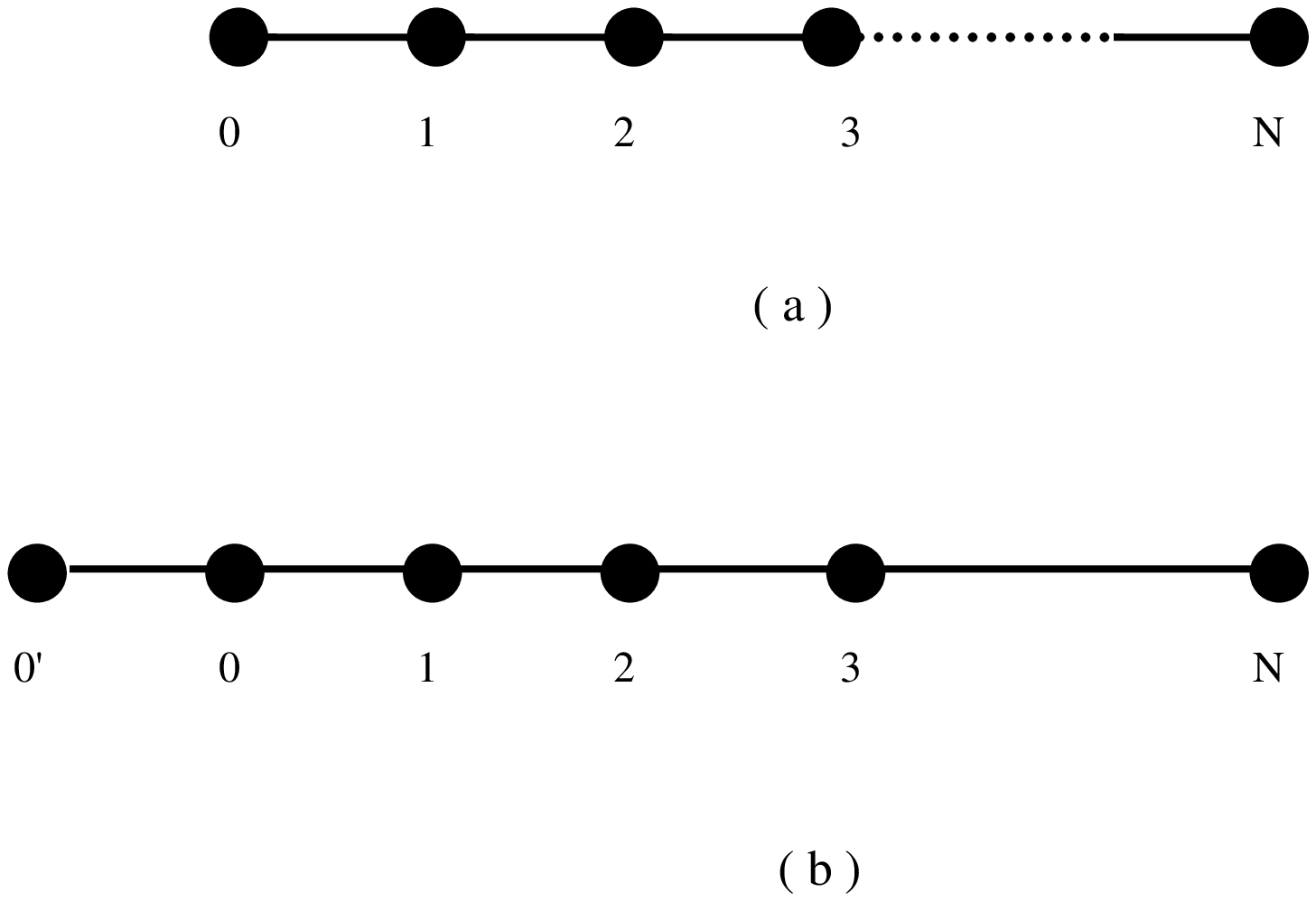}
    \caption{a-An unknown quantum state $|\phi\ra $ is placed at site $0$ of the chain
     and is transported to site $N$ by the dynamics of the spin
     chain.  b- The entanglement of a Bell state $|\phi^+\ra=\frac{1}{\sqrt{2}}(|0,0\ra+|1,1\ra)$
      placed at sites $0'$ and $0$
      develops into the entanglement of sites $0'$ and $N$.}
\end{figure}

The Hamiltonians governing the interaction of the spins in the
channel and the full chain are respectively
\begin{equation}\label{Heis}
    H_c = -J\sum_{i=1}^{N-1} {\bf \sigma}_i\cdot{\bf \sigma}_{i+1} +
    B \sum_{i=1}^{N} \sigma_{z,i},
\end{equation}
and
\begin{equation}\label{Heis2}
    H = -J\sum_{i=0}^{N-1} {\bf \sigma}_i\cdot{\bf \sigma}_{i+1} +
    B \sum_{i=0}^{N} \sigma_{z,i},
\end{equation}

 where the subscript "c" on $H$ stands for the channel.  If at
time $t=0$ the qubit $0$ is placed to the left of the chain then
the evolution of the Heisenberg chain carries the state of this
qubit to the rightmost spin $N$ where one extracts the state with
a rather high fidelity, provided that
one extracts the state at an optimal time.\\
It has been shown in \cite{bose} that one can transmit quantum
states with a high fidelity ranging from $F=1$ for $N=4$ to a
value exceeding $0.9$ for $N=7,10,11,13$ and $14$. The fidelity
generally decreases with increasing the length of the channel
exceeds the classical value
of $F=2/3$  for $N\approx 80$ which holds for classical transmission of quantum states\cite{horedecki}.\\
One can also use this channel for transmission of entanglement in
the following way \cite{bose}. One places two maximally entangled
spins labeled $0'$ and $0$ to the left of the chain (figure 1-b)
and evolution of the chain after a suitable lapse of time
entangles the spin $0'$ to the rightmost spin $N$ here it is
assumed that only the spin $0$ is coupled to the channel. In this
way two distant spins can be entangled which can later be used for
implementation of other
quantum protocols like teleportation \cite{tel}.\\

The formalism of \cite{bose} requires that the spins of the
channel be ideally aligned in the direction opposite to that of
the magnetic field. This ideal situation is however achievable
only at zero temperature or in very strong magnetic fields where
thermal fluctuations are not large enough to populate excited
states, i.e. (when $B/kT>> 1$). On the other hand increasing the
magnetic field may lower the quality of the channel since a high
magnetic field tends to align the spins and will generally
dominate the interaction between the spins
which is essential for the working of the channel. \\
Moreover when one uses this channel once and extracts the state at
the right hand site, the initial state of the channel turns into a
mixed state. Before using the channel for another round of
transmission the initial state of the channel should be restored,
for example by cooling to low enough temperatures. It is plausible
to assume that multiple uses of the channel may heat it up to
temperatures in which not only the ground state but also some of
the excited states are also populated. \\

In view of these considerations it is desirable to study the
effect of thermal fluctuations on such a quantum channel. This is
the problem that we want to address in this paper. Thus we want to
generalize the protocol of \cite{bose} to the case where the
initial state of the channel is not the ground state but a thermal
state given by a thermal density
matrix. \\
This enables us to to see the effect of the ambient temperature on
the feasibility of the protocol, and the quantitative effect that
temperature has on
the fidelity of transmission of states and distribution of entanglement.\\
We will derive a low temperature expansion through which we can
study the effect of temperature to any desired degree of accuracy
by
keeping appropriate number of terms in the expansion. \\
In a sequel to this paper we will study long chains of arbitrary
number of spins at low temperatures. This study can be done only
numerically. In this paper however we will study exactly thermal
effects on a short chain of four spins. The advantage of studying
this short chain is that we can obtain the spectrum completely and
hence can compare the two cases of ferromagnetic and
antiferromagnetic chains. \\

The structure of this paper is as follows: In section
{\ref{general}} we set up the general formalism and will develop
low temperature expansions for the expressions of fidelity and
entanglement. In section \ref{entanglement} we derive general
expressions for entanglement of endpoints of the chain at
arbitrary temperatures. In section \ref{example} we study exactly
the specific example of a short chain of only $4$ spins where we
present our basic results in figures (\ref{fidferro} to
\ref{conantiferro}).

\section{Low temperature expansion of the fidelity}\label{general}

We consider the interaction between the spins to be nearest
neighbor and of Heisenberg type. The spins also interact with an
external magnetic field. We should emphasize that much of what we
derive in this section do not depend on a specific form of the
Hamiltonian. We assume that the initial state of the channel is
\begin{equation}\label{thermal}
\rho_{th} = \frac{e^{-\b H_c}}{Z}=\sum_{\a}\frac{e^{-\b
E_{\a}}}{Z}|\a\ra\la \a|,
\end{equation}
where $|\a\ra$'s and $E_{\a}$'s are respectively the eigenstates
and eigenenergies of the channel Hamiltonian (\ref{Heis}). Here
$Z$ is the partition function of the
channel $Z:=tr(e^{-\b H_c})$.\\
 At time $t=0$ we place the $0$-th spin which is in an unknown
 pure
state $\rho_0=|\phi\ra \la \phi|$ to the left of this channel. The
initial state of the whole chain will be given by $\rho_0\otimes
\rho_{th}$ which evolves to
\begin{equation}\label{rho(t)}
\rho(t)= e^{-iHt}(\rho_0\otimes \rho_{th})e^{iHt},
\end{equation}
where $H$ is now the Hamiltonian of the full chain (\ref{Heis2}). \\
The state of the $N$-th spin after time $t$ will be given by
\begin{equation}\label{rho(N)}
\rho_N(t)= tr_{\hat{N}}(e^{-iHt}(\rho_0\otimes \rho_{th})e^{iHt}),
\end{equation}
where $\hat{N}$ means that we take the trace over all sites except
the $N$-th site. We can now derive an operator sum representation
for the transformation of the state of the leftmost qubit
$\rho(0):=|\phi\ra\la \phi|$ to the state of the rightmost qubit
$\rho_N(t)$. To do this we use (\ref{rho(N)}) to compute an
element of $\rho_N(t)$ as follows, where we use the index $I$ to
run over a complete set of states $\{|I\ra\}$ for the Hilbert
space of part of the total chain from site $0$ to site $N-1$, and
the index $\a$ is to run over the eigenstates of the channel
Hamiltonian:
\begin{eqnarray}\label{Kraus}
\la k|\rho_N(t)|l\ra &=& \sum_{I} \la I,k|e^{-iHt}(\rho_0\otimes
\rho_{th})e^{iHt}|I,l\ra\cr &=& \sum_{I,j,m,\a,\b} \la
I,k|e^{-iHt}|j,\a\ra\la j,\a|(\rho_0\otimes \rho_{th})|m,\b\ra \la
m,\b|e^{iHt}|I,l\ra\cr &=&\sum_{I,j,m,\a} \la I,k|e^{-iHt}|j,\a\ra
{\rho_0}_{j,m}\la m,\a|e^{iHt}|I,l\ra \frac{e^{-\b E_{\a}}}{Z}.
\end{eqnarray}
 Defining a collection of two by two matrices $M_{I,\a}$ with elements
 \begin{equation}\label{M}
\la k|M_{I,\a}|j\ra :=\frac{e^{\frac{-\b E_{\a}}{2}}}{\sqrt{Z}}\la
I,k|e^{-iHt}|j,\a\ra
 \end{equation}
we find the following Kraus decomposition \cite{kraus,Nieloson}
for $\rho_N(t)$,
\begin{equation}\label{Krausfinal}
    \rho_N(t)=\sum_{I,\a}M_{I\a}\rho_0M^{\dagger}_{I\a}.
\end{equation}

Note that in general the number of elements in this decomposition,
i.e. the number of matrices $M_{I,\a}$ is huge. In fact this
number is equal to the number of different choices for the pair of
indices $(I,\a)$, which equals $2^{2N}$. In principle there are
many operator sum representations for a superoperator and the
number of kraus operators for a qubit can be reduced to $4$
\cite{kraus,Nieloson}, however the present form has the advantage
that it is suitable for a low temperature expansion of the
fidelity of the channel. Note also that even in the present form
symmetry arguments highly restricts the number
of nonzero matrices as we will see in the sequel.\\

The fidelity of this out-state $\rho_N(t)$ with the in-state
$\rho_0$ is given by
\begin{equation}\label{F}
    F = \sum_{I,\a} tr(\rho_0M_{I\a}\rho_0M^{\dagger}_{I\a}).
\end{equation}
We are interested in the fidelity averaged over all the initial
input states, that is
\begin{equation}\label{F}
   \overline{F} = \frac{1}{4\pi} \int F d\Omega ,
   \end{equation}
where the integral is taken over the surface of the Bloch sphere.
This integral can further be simplified by using the following
easily verified identity
\begin{equation}\label{identity}
    tr(\rho_0 A \rho_0 B)=tr((A\otimes B)S(\rho_0\otimes \rho_0))
\end{equation}
where $S$ is the swap operator with elements
$S_{ij,kl}=\delta_{il}\delta_{jk}$. We now write $\rho_0$ as
$\rho_0=\frac{1}{2}({\bf 1} + {\bf n}\cdot {\sigma})$ and use the
identity $\frac{1}{4\pi}\int n_in_j d\Omega =
\frac{1}{3}\delta_{ij}$ to arrive at the following identity
\begin{equation}\label{average}
    \frac{1}{4\pi}\int \rho_0\otimes \rho_0d\Omega = \frac{1}{6}(\bf{1}+S),
\end{equation}
  which we use to rewrite $\overline{F}$ as

\begin{equation}\label{Fbar}
    \overline{F} =\frac{1}{6} \sum_{I,\a}
    tr(M_{I\a}\otimes M^{\dagger}_{I\a}(\bf{1}+S)).
\end{equation}

Using the facts that $tr((A\otimes B)S)= tr(AB)$ and $\sum_{I,\a}
M_{I\a}^{\dagger}M_{I\a}= \bf{1} $ we find the final form of
$\overline{F}$ as
\begin{equation}\label{Ffinal}
    \overline{F} = \frac{1}{3}+\frac{1}{6}\sum_{I,\a}
|tr M_{I\a}|^2.
\end{equation}
Equation (\ref{Ffinal}) is already in the form of a low
temperature expansion for the average fidelity. The leading
contribution comes from the ground state which we label as $\a_0$,
the next to leading contribution comes from the first excited
states and so on. Thus despite the huge number of matrices
$M_{I,\a}$, at low temperatures one can obtain a reasonably
good value of the fidelity by using only the first few terms in the expansion.\\
\subsection{The zero temperature limit of the ferromagnetic chain}
In this limit only the ground state contributes to the expansion
(\ref{Ffinal}). For the ferromagnetic chain the ground state is
\begin{equation}\label{gs}
    |\a_0\ra=|-,-,-,\cdots ,-\ra
\end{equation}
where all the spins are down. Thus we have
\begin{equation}\label{gs1}
\overline{F} = \frac{1}{3} + \frac{1}{6}\sum_{I} |tr M_{I\a_0}|^2.
\end{equation}
In (\ref{gs1}) it appears that a set of $2^N$ matrices
$M_{I,\a_0}$ contribute to the sum. However we show that by
symmetry considerations one can reduce this number to only two. To
this end we note that a general element of $M_{I,\a_0}$ can be
written as follows:
\begin{equation}\label{gs2}
    M_{I,\a_0}(i,j)=\la I,i|e^{-iHt}|j,-,-,-,\cdots ,-\ra .
\end{equation}
In view of the symmetry $[H,J_z]=0$ where $J_z$ is the total spin
in the $z$ direction, the only nonzero matrices are those in which
the index $I$ is either $(-,-,-,\cdots ,-)$ for which we denote
the corresponding matrix by $M_{0}$, or those
 in which only one spin is up (e.g. in the $i$-th position
 $(-,-,\cdots-, +,-,\cdots ,- )) $ for which we denote the corresponding matrix by
 $M_{i}$.\\
 Moreover the above mentioned symmetry requires that matrices be of the following form:

\begin{eqnarray}\label{Ms}
    M_0&=&\left(\begin{array}{cc}
         m_+ & 0 \\
                0 & m_-
              \end{array}\right),\cr
 M_i&=&\left(\begin{array}{cc}
         0 & 0 \\
                m_i & 0
              \end{array}\right) \h i = 1 \cdots N ,
\end{eqnarray}
where
\begin{eqnarray}\label{m+m-}
    m_+:&=& \la -,-,\cdots -,+|e^{-iHt}|+,-,-,\cdots ,-\ra,\cr
    m_-:&=& \la -,-,\cdots -,-|e^{-iHt}|-,-,-,\cdots ,-\ra,
\end{eqnarray}

and
\begin{equation}\label{mi}
    m_i:= \la -,\cdots -,+,-\cdots -|e^{-iHt}|+,-,-,\cdots ,-\ra .
\end{equation}

A simple calculation now shows that the following identity holds
for the above types of matrices, regardless of their explicit form
of matrix elements,
\begin{equation}\label{identity}
    \sum_{i=1}^{N} M_i\rho M_{i}^{\dagger} = \textbf{M}\rho
    \textbf{M}^{\dagger},
\end{equation}
where
\begin{equation}\label{mbig}
    \textbf{M} =\left(\begin{array}{cc}
        0 & 0 \\
                \sqrt{\sum_{i=1}^N|m_i|^2} & 0
              \end{array}\right)\\ .
\end{equation}
Thus the operator sum representation reduces to a sum with only
two elements, i.e.
\begin{equation}\label{2matrices}
    \rho_N(t) =M_0\rho_0M_0^{\dagger} +
    \textbf{M}\rho_0\textbf{M}^{\dagger}.
\end{equation}

Using (\ref{gs1}) and noting that the matrix ${\textbf{M}}$ is
traceless, we find the average fidelity at zero temperature:
\begin{equation}\label{FT=0}
    \overline{F} = \frac{1}{3}+\frac{1}{6}|m_+ + m_-|^2
\end{equation}
For the ferromagnetic chain, the state $|-,-,\cdots ,-\ra$ is also
the ground state of the full Hamiltonian and thus by shifting the
zero energy of the Hamiltonian to $0$ we can set $m_-=1$. Thus we
find
\begin{equation}\label{FT=1}
    \overline{F} = \frac{1}{3} + \frac{1}{6}|1 + m_+|^2
\end{equation}
in accordance with the result of \cite{bose}. Note that our $m_+$ is denoted by $f_{0,N}(t)$ in \cite{bose}. \\

\section{Transfer of entanglement}\label{entanglement}
We now consider how temperature affects the distribution of a
maximally entangled pair through the channel. Following
\cite{bose} we consider a maximally entangled pair of qubits in
the state $|\Phi_+\ra:= \frac{1}{\sqrt{2}}(|0,1\ra+|1,0\ra)$. In
figure (1-b) this pair of qubits are labeled by $0'$ and $0$.

The evolution of the chain may transform the entanglement between
the pair $(0',0)$ to the pair $(0',N)$, thus enabling us to
transport entanglement between ions or any other realization of
qubits over long distances. Note that the Hamiltonian only acts on
the part of the chain form $0$ to $N$. It is assumed that the
qubit $0'$ does not interact with the rest of the chain. After a
time $t$, the density matrix of the pair $(0',N)$ is easily
obtained thanks to the operator sum representation (\ref
{Krausfinal}). We find
\begin{equation}\label{ent1}
    \rho_{0',N}(t) =\sum_{I,\a} (\mathbf{1}\otimes M_{I\a})(|\Phi_+\ra\la\Phi_+)(\mathbf{1}\otimes
    M^{\dagger}_{I\a}).
\end{equation}
Using the fact that
\begin{equation}
|\phi_+\ra\la \phi_+|=\frac{1}{2} \left(\begin{array}{cccc}
   0&  &  &  \\
   & 1 & 1 &  \\
   & 1 & 1 &  \\
   &  &  &0  \\
  \end{array}\right)
\end{equation}
and the explicit form of the matrix elements of $M_{I,\a}$ as
given in (\ref{M}) we find after some manipulations the following
low temperature expansion:
\begin{equation}\label{ent}
    \rho_{0',N}(t):= \sum_{\a} \frac{e^{-\b
    E_{\a}}}{Z}\rho_{(\a)}(t),
\end{equation}
where each $\rho_{\a}$ pertains to a level $\a$ of the spectrum of
the channel and is given by
\begin{equation}\label{rhoa}
    \rho_{(\a)}(t)=\left(\begin{array}{cccc}
                  u^+_{\a}(t) &  &  &  \\
                            & w^+_{\a}(t) & z_{\a}(t) &  \\
                            & z^*_{\a}(t) & w^-_{\a}(t) &  \\
                            &  &  & u^-_{\a}(t) \\
                         \end{array}\right)
\end{equation}
where
\begin{eqnarray}\label{rhoaa}
u^+_{\a}(t)&=& \frac{1}{4}\la 1,\a| (1+\sigma^z_N(t))|1,\a\ra\cr
u^-_{\a}(t)&=& \frac{1}{4}\la 0,\a| (1-\sigma^z_N(t))|0,\a\ra\cr
w^+_{\a}(t)&=& \frac{1}{4}\la 1,\a| (1-\sigma^z_N(t))|1,\a\ra\cr
w^-_{\a}(t)&=& \frac{1}{4}\la 0,\a| (1+\sigma^z_N(t))|0,\a\ra\cr
z_{\a}(t)&=& \frac{1}{2}\la 0,\a| \sigma^+_N(t)|1,\a\ra.
\end{eqnarray}
Here the operators $\sigma^a_N(t)$ are operators in the Heisenberg
picture, i.e. ($\sigma^a_N(t)= e^{iHt}\sigma^a_Ne^{-iHt}$).\\
Thus the concurrence will have the same dependence on the
correlation function as in the thermal equilibrium state \cite{ok,
wz}, only now the correlation functions are time
dependent.\\

\section{Exact solution for a short chain}\label{example}
\begin{figure}
    \includegraphics[width=10cm,height=8cm,angle=0]{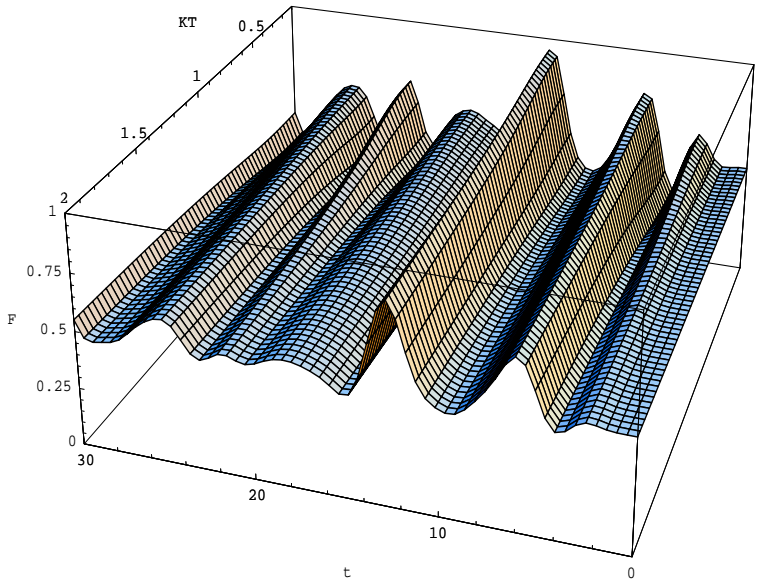}
    \caption{(Color Online) The fidelity (F) between the output and the input states averaged
    over all input states, as a function of $kT$ and time $t$, in a fixed magnetic field.
    In all the figures the fidelity $F$ and the concurrence $C$ are dimensionless and we are working in units where
 $KT$ and $t$ have no dimensions. In these units we have taken $B=1$ and $J=+1$ for ferromagnetic and $J=-1$ for anti-ferromagnetic chains.} \label{fidferro}\end{figure}

\begin{figure}
\centering
    \includegraphics[width=10cm,height=8cm,angle=0]{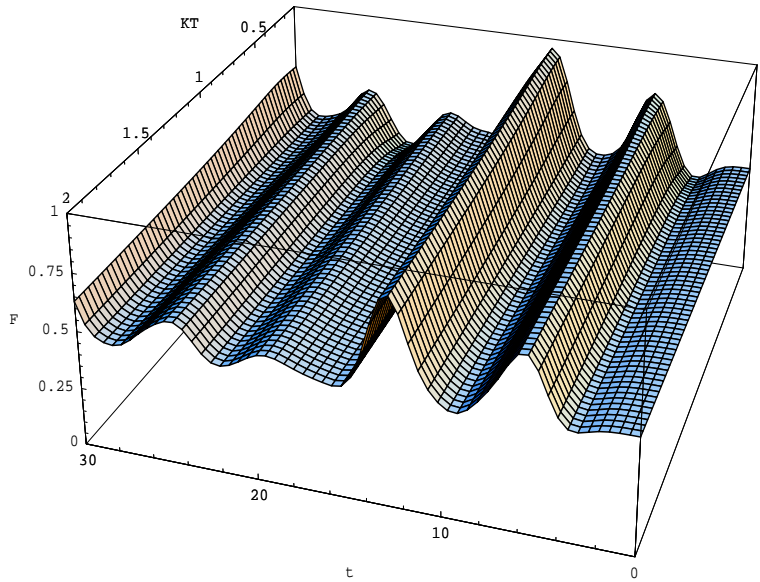}
    \caption{(Color Online) The fidelity (F) between the the output and the input states averaged
    over all input states, as a function of $kT$ and time (t), in a fixed magnetic field $(B=1)$,
    for an anti-ferromagnetic channel $(J=-1)$.}
    \label{fidantiferro}\end{figure}

An exact study of the thermal effects on a long chain with
arbitrary number of spins is highly involved, since it requires a
knowledge of all the energy eigenstates. On can study these chains
only at low temperatures in which case only the ground state and
the first excited states of the chain are populated. We will do
this in a sequel to this paper. Here we study exactly a short spin
chain of $N=3$.  An exact study of a short chain has the advantage
that one can compare the characteristically different behaviors of
ferromagnetic and anti-ferromagnetic chains. We remind the reader
that at zero temperature the channel relaxes to its ground state
which for the ferromagnetic chain is a disentangled state of spins
aligned with the magnetic field. This is the only case which has
been studied in \cite{bose}. \\ So we consider a three site
channel with hamiltonian given by
\begin{equation}\label{hc3}
    H_c = - J ({\bf s}_1\cdot {\bf s}_2+ {\bf s}_2\cdot {\bf s}_3)+
    B(s_{1z}+s_{2z}+s_{3z}).
\end{equation}
The spectrum of this hamiltonian is easily obtained and it exist
in the appendix.\\
 The Hamiltonian of the full chain is now
\begin{equation}\label{hc4}
    H = - J ({\bf s}_0\cdot {\bf s}_1+ {\bf s}_1\cdot {\bf s}_2+{\bf s}_2\cdot {\bf s}_3) +
    B(s_{0z}+s_{1z}+ s_{2z}+s_{3z}).
\end{equation}
Determination of the spectrum of this hamiltonian is facilitated
by using the following symmetries, namely
\begin{equation}\label{sym}
    [H,J_z]=[H,\Lambda]=[J_z,\Lambda]=0 \ \ \ {\rm and} \ \ \
    \sigma_x^{\otimes 4} H(J,B) = H(J,-B)  \sigma_x^{\otimes 4},
\end{equation}

where $\Lambda$ is the inversion operator
\begin{equation}\label{inv}
    \Lambda |s_0,s_1,s_2,s_3\ra = |s_3,s_2,s_1,s_0\ra.
\end{equation}
The spectrums of the total and the channel hamiltonian are derived
in the appendix. By plugging these eigenstates and eigenvalues in
equations (\ref{M}) and (\ref{Ffinal}) one can determine the
average fidelity of the output state with the input state.  The
fidelity is a complicated function of time, in fact it is a
superposition of periodic functions with periods
$\omega_{ij}:=\frac{1}{|E_j-E_i|}$ . In \cite{bose} one extracts
the output state only at certain times where the fidelity reaches
a maximum. Since we want to focus on the effect of temperature, we
fix the magnetic field and determine the average fidelity as a
function of time and temperature. The results are shown in figure
(\ref{fidferro}) and (\ref{fidantiferro}) for ferromagnetic and
anti-ferromagnetic chains respectively.

\begin{figure}
\centering
    \includegraphics[width=10cm,height=8cm,angle=0]{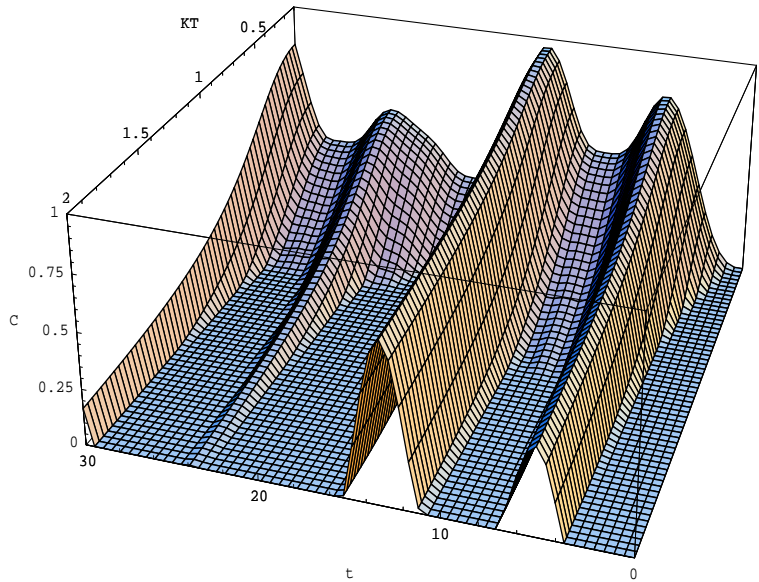}
    \caption{(Color Online) The concurrence $C$ of the state of the endpoints of the chain as a function of time $t$ and
    $kT$ in a fixed magnetic field for a ferromagnetic chain.  Here $B=1$ and $J=1$.
    }
   \label{conferro}\end{figure}

\begin{figure}
\centering
    \includegraphics[width=10cm,height=8cm,angle=0]{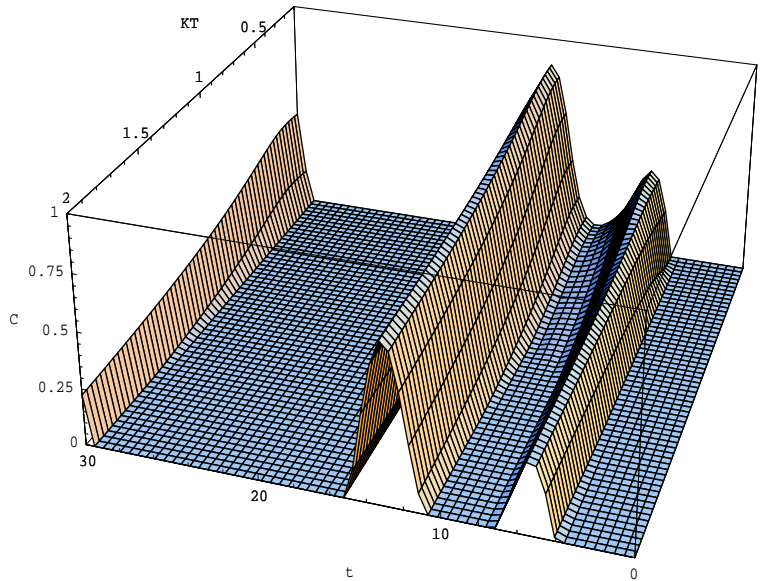}
    \caption{(Color Online)The concurrence $C$ of the state of the endpoints of the chain as a function of time $t$ and
    $kT$, in a fixed magnetic field $(B=1)$, for a anti-ferromagnetic chain $(J=-1)$.}
\label{conantiferro}\end{figure}

These curves show several interesting features. The first one is
that the optimal time of extraction is almost independent of
temperature, thus at any temperature one can tune the optimal time
of extraction to be the same as that of the zero temperature.  The
only effect of temperature is that it decreases the fidelity. It
is also seen that the optimal times when the fidelity reaches
local maxima are the same for both types of chains. Moreover
thermal fluctuations have much less destructive effect on the
fidelity in
the anti-ferromagnetic chain as compared with the ferromagnetic chain.\\
The difference between these two types of chains is more
pronounced when we use them to transfer entanglement. We have used
these two channels to transfer the maximal entanglement between
the two spins $0'$ and $0$ at the left hand side of the chain
(\ref{mychain2}) to entanglement of the end points of the chains
at time $t$, measured by the concurrence of the density matrix
$\rho_{0',3}$. In a fixed magnetic field, the concurrence of this
density matrix \cite{woo} is a function of temperature. Figures
(\ref{conferro}) and (\ref{conantiferro}) show this concurrence as
a function of temperature and time for the ferromagnetic and
anti-ferromagnetic chains respectively. All the previous comments
apply also to this type of behavior. The striking difference is
that in the anti-ferromagnetic chain, there are long intervals of
time when no entanglement can be distributed in the chain
regardless of the temperature, entanglement transfer is possible
only in short periods of time. In fact comparison of the figures
for the fidelity and concurrence shows that when the fidelity of
the channel drops below the approximate value of $0.6$, it no
longer can transfer any entanglement.

\section{Summary}
We have studied the effect of thermal fluctuations on a recently
proposed method for transportation of unknown states through
quantum spin chains. We have developed a low temperature expansion
which can be used to calculate this effect to a desired degree of
accuracy at any given temperature. As an example we have
calculated exactly the effect of thermal fluctuations on
transportation of states on a short spin chain and have shown that
the optimal time of extraction of transported states at the end of
the chain is almost independent of temperature, the only effect of
which is to lower slightly the fidelity of the output state with
the input state. We have made a detailed comparison between the
ferromagnetic and anti-ferromagnetic channels.
\section{Acknowledgement}
A. Bayat would like to thank A. T. Rezakhani for his help in
drawing the figures. We also thank I. Marvian for his very
constructive comments and M. Asoudeh and L. Memarzadeh for their
critical reading of the manuscript.
\section{Appendix :Spectrum of three and four Site Spin chain }

In this appendix we collect the eigenstates and eigenvalues of the
hamiltonians $H_c$ and $H$ shown in equations (\ref{hc3}) and
(\ref{hc4}).\\  The eigenstates and eigenenergies of the channel
are as follows:
\begin{eqnarray}\label{spec3}
    |\a_1\ra &=& |-,-,-\ra,\cr
|\a_2\ra &=& \frac{1}{2}(|+,-,-\ra - \sqrt{2}|-,+,-\ra
+|-,-,+\ra,\cr |\a_3\ra &=& \frac{1}{2}(|+,-,-\ra +
\sqrt{2}|-,+,-\ra +|-,-,+\ra,\cr
 |\a_4\ra &=& \frac{1}{\sqrt{2}}(|+,-,-\ra -
|-,-,+\ra),
\end{eqnarray}
and
\begin{equation}\label{spec3i}
    |\a_{i+4}\ra = \sigma_x^{\otimes 3}|\a_i\ra  \h i = 1,
    \cdots ,4,
\end{equation}
with energies
\begin{eqnarray}\label{ener3}
E_1 &=& -\frac{J}{2}-\frac{3B}{2},\cr
    E_2 &=& -\frac{J}{2}-\frac{B}{2},\cr
    E_3 &=& J-\frac{B}{2},\cr
    E_4 &=& -\frac{B}{2},
\end{eqnarray}
and
\begin{equation}\label{ener3i}
    E_{i+4}(J,B) = E_i(J,-B)  \h i = 1,
    \cdots ,4.
\end{equation}
The eigenstates of the total hamiltonian H (equation (\ref{hc4}))
are obtained by using the symmetries (\ref{sym}).
 We use the notation
$|i\ra$ or $|i,j\ra$ to indicate that the spins in the $i-$ th
position or the $(i,j)$ positions are up and the rest are down:

\begin{eqnarray}\label{}
|\chi_{1}\ra &=& |-,-,-,-\ra\cr |\chi_2\ra &=& \frac{1}{2}(|1\ra
+|2\ra + |3\ra+|4\ra)\cr
 |\chi_{3}\ra &=&\frac{1}{2}(|1\ra-|2\ra-
|3\ra+|4\ra)\cr |\chi_{4}\ra &=&
\frac{1}{2\sqrt{2+\sqrt{2}}}(|1\ra-(\sqrt{2}+1)(|2\ra
-|3\ra)-|4\ra)\cr |\chi_{5}\ra &=&
\frac{1}{2\sqrt{2-\sqrt{2}}}(|1\ra+(\sqrt{2}-1)(|2\ra
-|3\ra)-|4\ra)\cr
 |\chi_6\ra &=&\frac{1}{\sqrt{2}}(|1,4\ra - |2,3\ra)\cr
 |\chi_{7}\ra &=&
\frac{1}{\sqrt{6}}(|1,2\ra + |1,3\ra + |1,4\ra + |2,3\ra + |2,4\ra
+ |3,4\ra)\cr |\chi_8\ra &=&\frac{1}{2\sqrt{(2+\sqrt{2})}}(|1,2\ra
- (1+\sqrt{2})(|1,3\ra -|2,4\ra)-|3,4\ra)\cr
 |\chi_9\ra
&=&\frac{1}{2\sqrt{(2-\sqrt{2})}}(-|1,2\ra + (1-\sqrt{2})(|1,3\ra
- |2,4\ra)+|3,4\ra)\cr
 |\chi_{10}\ra &=&
\frac{1}{2\sqrt{3\eta_+}}(|1,2\ra -\eta_+|1,3\ra + \xi_+|1,4\ra+
\xi_+|2,3\ra - \eta_+|2,4\ra + |3,4\ra)\cr
 |\chi_{11}\ra &=&
\frac{1}{2\sqrt{3\eta_-}}(|1,2\ra -\eta_-|1,3\ra + \xi_+|1,4\ra+
\xi_-|2,3\ra+ -\eta_-|2,4\ra + |3,4\ra)
\end{eqnarray}
where $\xi_{\pm} := 1\pm \sqrt{3}$ and $\eta_{\pm} = 2 \pm
\sqrt{3}$.  The other five states are obtained by the action of
the flip operator ${\sigma_x}^{\otimes 4}$ on the first five
states above, that is:
\begin{equation}\label{rest}
   |\chi_{i+11}\ra =\sigma_x^{\otimes 4}|\chi_i\ra \h i = 1, \cdots,
   5.
\end{equation}
The energies of the above states are:
\begin{eqnarray}\label{ener4}
    E_1 &=& - \frac{3}{4}J - 2B \h\ \ \ \ \ \ \ \  E_2 = - \frac{3}{4}J - B \h \ \ \ \ \ \ \ E_3 =  \frac{1}{4}J - B \cr
    E_4 &=& \frac{1}{4}(1+2\sqrt{2})J -B \ \ \ \ \ \ \  E_5 =
\frac{1}{4}(1-2\sqrt{2})J - B \ \ \ \ E_6 = \frac{1}{4} J \cr E_7
&=& - \frac{3}{4}J  \h\h\h E_8 = \frac{1}{4}(1+2\sqrt{2})J \ \ \
 \h E_9 = \frac{1}{4}(1-2\sqrt{2})J \cr \ \ \ E_{10} &=&
\frac{\sqrt{3}}{4}\eta_+ J \ \ \ \h\h  E_{11} =
-\frac{\sqrt{3}}{4}\eta_- J.
\end{eqnarray}
and
\begin{equation}\label{x}
    E_{i+11}(J,B) = E_{i}(J,-B)\h i=1,\cdots ,5 .
    \end{equation}

\end{document}